# Interband plasmon polaritons in magnetized charge-neutral graphene

T. M. Slipchenko[1,2*], J.-M. Poumirol[3,4], A. B. Kuzmenko[3*], A. Yu. Nikitin[5, 6], L. Martín-Moreno[1,2*]

[1] Instituto de Nanociencia y Materiales de Aragón (INMA)

CSIC-Universidad de Zaragoza, Zaragoza 50009, Spain

[2] Departamento de Física de la Materia Condensada,

Universidad de Zaragoza, Zaragoza 50009, Spain

[3] Department of Quantum Matter Physics, University of Geneva,

[4] CEMES, University of Toulouse, CNRS, Toulouse 31000, France

[5] IKERBASQUE Basque Foundation for Science, 48013 Bilbao, Spain.

[6] Donostia International Physics Center (DIPC), 20018 Donostia-San Sebastin, Spain

**ABSTRACT:** Studying the collective excitations in charge neutral graphene (CNG) has recently attracted a great interest because of unusual mechanisms of the charge carrier dynamics. The latter can play a crucial role, for instance, for superconducting phases in the periodically strained CNG and the magic angle twisted bilayer graphene or for formation of graphene plasmon polaritons (GPPs) associated with interband transitions due to the strain-induced pseudomagnetic field. Importantly, GPP in CNG can be a tool providing new insights into various intriguing quantum phenomena in CNG via optical experiments. However, interband GPPs in CNG are barely investigated, even in the simplest configurations. Here, we show that magnetically biased single layer CNG (particularly, at zero temperature) can support interband GPPs of both transverse magnetic and transverse electric polarizations. They exist inside the narrow absorption bands originating from the electronic transitions between Landau levels and are tunable by the external magnetic field. We put our study into the context of potential near-field and far-field optical experiments, thus opening the door to the exploration of CNG for optics and nanophotonics.

## Introduction

GPPs - electromagnetic fields coupled to oscillating Dirac charge carriers in graphene - exhibit nanoscale field confinement allowing for an active control of their wavelength via gating, and are thus highly interesting for nanophotonics and optoelectronics [1]. To support GPPs, graphene needs to host a sufficient amount of free charge carriers, which can be achieved by doping (chemical or via gating) [2, 3], thermal fluctuations [4, 5] or optical pumping [6]. In doped graphene, currently considered as a standard platform for exploring GPPs, the energies of GPPs (and thus frequencies at which they can be excited) are limited by the Fermi energy, $E_F$, which can reach only a few tenths of an electronvolt. Therefore, intense efforts are currently underway to exploit other ways to tune GPPs at higher energies [7]. One way to tune GPPs in doped graphene is to apply a static magnetic field, resulting in so called graphene magnetoplasmons [8, 9, 28]. However, the performance of the magnetoplasmons in doped graphene is still limited by the high electron scattering rates and the need of high magnetic fields (typically, several Tesla).

Recently, it has been realized that in certain conditions CNG could become an alternative to doped graphene for GPP applications, since CNG possesses specific optical properties. For instance, being exposed to reasonably low magnetic fields, CNG manifests the colossal magneto-optical activity in the mid-infrared and terahertz frequency ranges [10]. Besides, as reported in [11], CNG shows the giant intrinsic photoresponse due to the small electron-electron scattering rates. Most importantly, collective effects in CNG (in particularly, charge neutral twisted bilayer) can enable the formation of GPP associated with the interband transitions [12, 13]. Intriguingly, these interband GPP can be closely related with the superconducting states appearing in periodically strained single layer CNG [14] and magic angle twisted bilayer graphene [15-17]. Up to now, although having a high interest for superconducting states formation and increasing of operational frequencies for graphene-based plasmonic devices, GPPs in CNG have been barely explored.

Here, motivated by the recent experimental data [10], we show that a single-layer CNG biased by external static magnetic field, or, simply, magnetized CNG (MCNG), can support interband GPPs of both transverse magnetic (TM) and transverse electric (TE) polarizations. Their wavelength and energy can be controlled by varying the amplitude of the applied magnetic field, while their lifetimes are comparable with low-loss phonon-polaritons recently discovered in van der Waals crystal slabs [18,19]. Remarkably, the interband TE GPPs have two orders of magnitude larger confinement of the electromagnetic fields than TE intraband GPPs [20-22]. A much better confinement of the fields in the interband TE GPPs favors their excitation via Bragg resonances in a graphene grating, as we demonstrate in our calculations. On the other hand, we show that TM GPPs can be very efficiently excited either via Fabry-Perot resonances in an array of graphene ribbons (in the far-field experiments) or by means of near-field microscopy (in the near-filed experiments).

**Results**

**The origin of interband GPPs in MCNG and their properties.**

Recall that the conventional (intraband) GPPs refer to the intraband electronic transitions (see Fig. 1a), which are possible in the doped graphene [23-27]. In contrast, in CNG the intraband electronic transitions are not allowed and thus GPPs cannot be supported. On the other hand, in MCNG, instead of the intraband transitions, the interband transitions between oppositely signed Landau levels (LLs) or the ones involving zero LL are allowed (see Fig. 1d). The electrons transiting between conduction and valence bands, can couple to the time-periodic electromagnetic field and build up an oscillating charge density which can be considered as GPP in MCNG (note that similar collective oscillations occur in doped graphene in magnetic field [28]). The properties of GPPs in MCNG are fully determined by the magneto-optical conductivity of MCNG, $\hat{\sigma}$. Although in general $\hat{\sigma}$ of a magnetized graphene is a tensor, in case of MCNG its diagonal components are identical, $\sigma_{xx} = \sigma_{yy} = \sigma$, and the off-diagonal conductivity vanish, $\sigma_{xy} = \sigma_{yx} = 0$. Thus, similarly to non-magnetized doped graphene the magneto-optical conductivity of MCNG is characterized by a scalar, $\sigma$, which can be calculated within the linear-response approximation [10, 29].

For simplicity and without loss of generality, we restrict our analysis of the GPPs dispersion by the free standing MCNG in vacuum. However, later on, in order to mimic realistic near-field and far-field experiments, we consider MCNG encapsulated between thin hBN slabs. Encapsulated graphene samples are commonly used in the optical experiments, as they show record-high mobilities of charge-carriers, and thus the highest GPPs lifetimes [30]. In addition, through the whole paper we will limit ourselves to CNG at zero temperature, T = 0. As we show in Supplementary Note 1, the dispersions of studied GPP in MCNG undergo negligible changes with an increase of both temperature, $k_B T$, and Fermi energy, $E_F$, up to $0.2\, E_c(B)$, where $E_c(B) = \sqrt{2e\hbar v_F^2 B}$ is the cyclotron energy, $v_F$ is Fermi velocity, $B$ is the magnetic field. For example, simultaneously rising $E_F$ and $k_B T$ from zero up to $0.2 E_c(B)$ at frequency 17 THz the dispersion shifts only by $\Delta k / k_{pl} \approx 0.006$, where $k_{pl}$ is wavevector of GPP in MCNG. For definiteness and unless explicitly stated, in all calculations we will consider that the relaxation time of the charge carriers in graphene is $\tau = 1\, ps$ (achievable in hBN encapsulated graphene [10]) and the Fermi velocity is $v_F = 1.15 \times 10^6 ms^{-1}$; while the external magnetic field is $B = 1.3\, T$ ($E_c = 0.048\, eV$) (reachable with neodymium magnets [31]).

In Fig. 1b and Fig. 1e we show the real and imaginary parts of the adimensional conductivity, $\alpha = 2\pi\, \sigma/c$, as a function of frequency, $\nu$, for non-magnetized doped graphene (with $E_F = 0.1\, eV$) and MCNG, respectively. Both Re($\alpha$) and Im($\alpha$) of MCNG behave in a markedly different manner compared to those of the non-magnetized doped graphene. Namely, in MCNG Re($\alpha$) has a series of pronounced peaks. These peaks correspond to the interband transitions, $|n-1| \to |n|$ and $|n| \to |n-1|$, occurring at the discrete frequencies $\nu_n = E_c \left( \sqrt{|n-1|} + \sqrt{|n|} \right)/h$. At these frequencies Im($\alpha$) changes its sign from negative to positive.

The type of the GPP mode is determined by the sign of Im($\alpha$). Indeed, the dispersion relation of TM and TE GPP reads [32]: $Y^\gamma + \alpha = 0$, where $\gamma$ stands for either TM or TE polarization, $Y^{TM} = 1/q_z$, $Y^{TE} = q_z$ and $q_z$ is the dimensionless out-of-plane wavevector component. Thus, the positive (negative) values of Im($\alpha$) guarantee the exponential decay of the GPP fields (depending upon z as $e^{iq_z k_0 |z|}$, where $k_0$ is wavevector in vacuum), of TM (TE) polarization.

In Fig. 1c, 1f we illustrate the dispersion curves for the GPPs in non-magnetized doped graphene (with $E_F = 0.1\ eV$ taken for reference) and MCNG, respectively. In each case, the dispersion of TM (TE) GPPs is shown by the blue (red) curves. We see that GPPs in MCNG and in non-magnetized doped graphene behave very differently. Namely, while in the non-magnetized doped graphene both TM and TE intraband GPPs exist in two continuous frequency intervals, in case of MCNG, the dispersion of the interband GPPs is split into a set of narrow frequency bands with the energy band width $h\Delta\nu \approx E_c$ for the first TM and TE bands. The band widths increase with both magnetic field and band index, $n$.

At the high-frequency border of the GPPs frequency bands, the dispersion curves of both TM and TE modes present a back-bending towards $q_{pl} = k_{pl}/k_0 = 0$, taking place due to the losses in graphene (given by Re($\alpha$)). Although the losses in MCNG inevitably limit the propagation of the GPPs, their figure of merit – defined as the ratio between GPP propagation length, $L_{pl}$, and the GPP wavelength, $\lambda_{pl}$, FOM = $L_{pl}/\lambda_{pl}$ – can be comparably large.

Indeed, in case of TM GPP in MCNG, for small values of α, we can approximate $\lambda_{pl} = \lambda\ \text{Im}(\alpha)$ and $L_{pl} = \frac{\lambda\ \text{Im}(\alpha)^2}{2\pi\ \text{Re}(\alpha)}$ so that FOM = $\text{Im}(\alpha)/2\pi\text{Re}(\alpha)$. The condition $Im(\alpha) \gg Re(\alpha)$, providing high FOM, is fulfilled at the frequencies in the middle of each TM band. Inside each band FOM grows proportionally to both the relaxation time and the magnetic field as $\propto (\tilde{v}_n(B) - v_n(B))\tau$, where both $\tilde{v}_n(B)$ and $v_n(B)$ are proportional to $\sqrt{B}$. For the realistic parameters considered in Fig. 1f, the FOM of TM GPP in the first frequency band reaches approximately 3.5 (see Supplementary Note 2). Thus, remarkably, FOMs of TM GPPs in MCNG can be of the same order of magnitude as FOM of phonon polariton in h-BN, [33]. This result is not intuitive since GPPs present completely different loss channels compared to phonon polaritons (electron-electron and electron-phonon scattering versus phonon-phonon scattering, respectively), but it may be very useful as, in contrary to the case of phonon polaritons, GPPs in MCNG are tunable.

To better illustrate the strong confinement of TM GPPs in MCNG in Fig. 2a we show the spatial distribution of the vertical electric field generated by a vertical electrical dipole positioned at $z = 60\ nm$ above the graphene, assuming the applied magnetic field of 1.3 T, and $\nu = 12\ THz$ ($\lambda_0 = 25\ \mu m$). The fringes of the opposite polarities propagating away from the dipole region clearly indicate the polaritonic wavelength ($\lambda_{pl}$= 4.3 $\mu m$) being much smaller than the free-space wavelength, $\lambda_0 = 25\ \mu m$, indicated by the horizontal black arrow (the corresponding frequency, $\nu = 12\ THz$, is marked by the black point in the red dispersion curve and by the horizontal dashed line in Fig.2d).

The properties of TM GPP in MCNGs strongly depend on the applied magnetic field. In Fig. 2b we plot the real part of the z-component of electric field of TM GPP, excited by the dipole, as a function of the distance along graphene, x, at two different values of the magnetic field: $B = 1.3\,T$ and $B = 1\,T$. These field profiles illustrate that TM GPP wavelength increases with B ($\lambda_{pl}$ is almost 5 times larger for $B = 1.3\,T$ than for $B = 1\,T$), the latter tendency of $\lambda_{pl}(B)$ being further confirmed by the red curve in Fig. 2c. The explicit dependence $\lambda_{pl}(B)$ can be estimated from the dispersion relation, shown in Fig. 2d for several values of the magnetic field within the first band. Away from the band-bending, the GPP wavelength is given by $\lambda_{pl} = \frac{\lambda_0}{q_{pl}} \approx \frac{2W_n}{\pi(\nu^2 - \nu_n^2)}$ (with $W_n = \frac{\sigma_{uni}}{\pi} \frac{E_c^2}{\hbar^2 \nu_n}$ being the spectral weight [10] and $\sigma_{uni} = \frac{e^2}{4\hbar}$ the universal optical conductivity), thus non-linearly scaling with B.

In strong contrast to the TM modes, the TE GPP in MCNG are far less confined, as shown by the magnetic field snapshot in Fig. 3a (where we use a vertical magnetic dipole source positioned at $z = 5\,\mu m$ above the graphene at the frequency of $\nu = 11.2\,THz$). The excitation of TE GPP in MCNG is confirmed by the diffraction shadow [34] seen in Fig. 3a. It appears because of the destructive interference of the excited TE GPP in MCNG and the dipole radiation. This is due to the comparable wavelengths of free-scape radiation and TE GPPs. At a sufficient distance from the source the TE plasmon is clearly distinguishable from the free-space fields. For a more quantitative analysis of the TE plasmons launched by the dipole, in Fig.3b we plot the field profile along graphene for two values of the magnetic field: $B = 1.3\,T$ (red curve) and $B = 1.4\,T$ (blue curve). For comparison, the field launched by the dipole in free-space without graphene (with the wavelength $\lambda_0 = 26.77\,\mu m$) is also shown by the black curve. According to the field profiles, the difference between $\lambda_0$ and the wavelength of TE GPP in MCNG does not exceed 2%. Besides, we see that, in contrast to TM modes, the wavelength of TE GPP in MCNG weakly depends on the applied magnetic fields, see Fig. 3c (red curve). Nevertheless, as the plasmons propagate away from the source, the phase shift between the field profiles for different magnetic fields becomes considerable (see the shift between the red and green field profiles in Fig 3b). The propagation length of TE GPP in MCNG is also much more sensitive to the magnetic field than their wavelength, as shown in Fig. 3c (in Fig 3b the red field profile clearly decays much quicker than the green one). The latter can be explained by the vertical shift of the dispersion curves of the TE plasmons with the increase of the magnetic field, and thus the back-bending region (where the absorption is increased), see Fig. 3d. The sensitivity of both the phase shift and the propagation length to the magnetic field can be potentially used for the applications of TE GPPs, particularly in interferometers.

Remarkably, TE GPPs in MCNG are much more confined to the graphene sheet than conventional TE GPPs. The strongest confinement takes place close to the back-bending of the dispersion curve, where the momentum reaches its maximum values $q_{pl} = q_{pl}^{(max)} \approx 1 + \frac{1}{2}\left[\frac{W_n \tau}{c}\right]^2$ at the frequencies $\nu^{(max)} \approx \nu_n - 2/(3\pi\tau)$, with, $\sim 1/q_{pl}^{(max)}$, becoming two orders of magnitude smaller than in non-magnetized doped graphene [15-17]. At lower frequencies TE GPP dispersion curve asymptotically tends to the light line, $q_{pl} = 1$ and thus becomes undistinguishable from the dispersion of the free-space waves.

GPPs in MCNGs can be potentially observed experimentally using either near-field microscopy of non-structured MCNG or far-field transmission/reflection spectroscopy of periodically patterned MCNG. The near-field microscopy, however, is more appropriate for TM GPP in MCNG, since the tip of the near-field microscope typically presents a vertically polarized dipole source, thus better matching with the TM polarized excitations. In the next sections we mimic both near-field and far-field experiments.

**Prospects for potential spectral near-field experiments for the characterization of TM GPPs in MCNG.**

Scattering-type near-field optical microscopy (s-SNOM) utilizes a tip of an atomic force microscope (AFM), which is illuminated with an external infrared laser [35]. The laser beam is focused at the apex of the tip, providing the necessary momentum to launch GPPs in graphene, as illustrated in Fig. 4(a). GPPs emanate from the tip and upon reaching the sample edge, they are reflected back. As the tip is scanned towards the edge, the back-scattered signal is collected in the detector as a function of the tip position [2, 3]. By using a broad-band light source, the near-field light scans can be represented as a function of frequency [6]. By Fourier transforming the interferogram as a function of frequency, the dispersion relation of GPPs can be retrieved.

In our numerical model we represent the illuminated AFM tip by a vertical dipole source. As has been shown in [36], the vertical component of electric field below the dipole, $|E_z|$, can serve as a good qualitative approximation for the signal scattered by tip. Therefore, we calculate the field below the dipole, $|E_z|$, as a function of the dipole position ($x$) and frequency ($\nu$), composing a near-field hyperspectral image, $|E_z(x, \nu)|$.

An example of the near-field hyperspectral image of the single layer of MCNG encapsulated between two flakes of hBN (the thicknesses of the top and bottom hBN layers are 5 nm and $10 \, nm$, respectively) is shown in Figure 4b. We perform simulations at the frequencies of the first frequency band of TM GPP in MCNG: $11.5 - 22.35 \, THz$.

In Fig. 4b we clearly observe several bright fringes (composed of the alternating NF minima and maxima) indicating the interference between the GPP launched by the dipole and GPP reflected by the edge. The distance between neighboring NF minima (or maxima) far from the edge equals approximately to the half of the GPP wavelength, $\lambda_{pl}/2$. As $\nu$ increases the fringes become thinner and their inter-spacing decreases. The latter decrease is consistent with the dependence $\lambda_{pl} \propto (\nu^2 - \nu_n^2)^{-1}$, where for the first band $\nu_1 = 11.5\ THz$. Besides, the amplitude of the fringes decays with the increase of $\nu$, which can be attributed to the increase of MCNG losses (i.e. growth of $Re(\alpha)$) particularly, when approaching the limiting frequency, $\tilde{\nu}_1 = 22.35\ THz$.

In Fig. 4c we plot the Fourier transform of the near-field hyperspectral image represented in Fig. 4b. The bright maximum seen in the color plot perfectly matches the GPP dispersion curve (the dashed green curve), assuming that the momenta of the GPPs is doubled, $2\ q_{pl}(\nu)$. The latter is consistent with the $\lambda_{pl}/2$ distance between the interference fringes in Fig. 4b. With our analysis we conclude that the GPPs in hBN-encapsulated MCNG should be observable in s-SNOM experiments for realistic parameters of graphene, even at moderate external magnetic fields.

**Prospects for far-field spectroscopy experiments.**

For the coupling with GPPs from the far-field, the graphene can be structured into ribbons [37-40]. Such structuring allows one to avoid the momentum mismatch between the GPPs and free-space waves. Depending upon the parameters of the grating, the excited GPPs present either "quantized" Fabry-Perot modes inside the ribbons, or Bragg-resonances manifesting themselves a dip/peak in the transmission/reflection spectrum spectra. In our simulations for the excitation of both TM and TE GPP in MCNGs we consider a periodic array of micro-ribbons made in either free-standing or hBN-encapsulated graphene, as illustrated in Fig. 5a.

In Fig. 5b we show the absorption spectra, $A(\nu)$, of the ribbon arrays (of different ribbon widths, $W$, and a fixed period, $L = 0.22\ \mu m$) illuminated by a normally-incident monochromatic plane wave. The latter is polarized along the *x*-direction, thus matching with the polarization of TM GPPs. The solid and dashed curves in Fig. 5b correspond to the absorption by the hBN encapsulated and free-standing MCNG ribbon arrays, respectively. In continuous MCNG (solid black curve in Fig. 5b), the absorption spectra present- only one maximum, matching with the interband transition frequency $\nu_1 = 11.5\ THz$. In contrast, the absorption spectrum of the MCNG ribbon array, (both for hBN-encapsulated and free-standing graphene), has a set of resonant maxima. For instance, at $W = 3L/4$ the absorption spectra (solid blue curve) has one strong resonance at $12.5\ THz$ and set of much weaker ones at higher frequencies. With decreasing $W$, the resonances redshift away from $\nu_1$ dropping in their magnitudes. Similarly to doped graphene ribbon arrays, the emergence of the absorption maxima can be explained by Fabry-Perot resonances in a single ribbon, forming while the GPPs reflect back and forth from the ribbon edges. In the resonance, the GPP refractive index, $q_{pl} = \frac{\lambda_0}{\lambda_{pl}}$, should satisfy [41]

$$q_{pl} = \frac{\lambda_0(m+3/4)}{2W}, \qquad (1)$$

where $m$ is an even number. Note that the modes with the odd integers have antisymmetric distribution of the vertical electric field with respect to the ribbon axis and do not couple to normally-incident waves. For all $W$, the absorption resonances in the spectrum of hBN-encapsulated graphene ribbon array appear at higher frequencies compared to the ones in the free-standing array. This can be explained by the stronger confinement of GPPs in encapsulated graphene than in the free-standing one. The strong confinement of TM GPPs makes them excellent candidates for sensing the environment (see, for example, [42]). In fact, changes in the environment can be detected via the resulting frequency shifts of the plasmonic resonances in the far-field spectrum. The absorption resonances are corroborated by the dispersion curves in the bottom panel of Fig. 5b (GPPs in the encapsulated graphene have larger momenta). The vertical dotted lines connecting the absorption maxima (Fig. 5b, top panel) with the GPP dispersion curves (Fig. 5b, bottom panel) mark the frequency positions of the peaks. The frequencies of the peaks match well with the ones given by the simple Fabry-Perot 0th-order resonances in Eq.(1). Substituting into Eq. (1) the approximate expression for the TM GPP wavelength found in the previous section, the frequency positions of the resonances can be explicitly written as

$$\nu \approx \sqrt{\frac{w_n}{\pi W}(n + 3/4) + \nu_n^2} \propto \sqrt{\frac{B}{W}} \qquad (2)$$

The resonance frequency can thus be tuned by both magnetic field and the ribbon width.

To resonantly excite the TE GPPs, we now illuminate the hBN-encapsulated graphene ribbon arrays by a normally-incident monochromatic plane wave with electric field pointing along the $y$ direction. Due to much smaller momenta of TE GPPs, the array period must be nearly two orders of magnitude larger than the one considered for excitation of TM plasmons. In Fig. 5c we present $A(\nu)$, as before, for different $W$ and for the same period $L = 28~\mu m$. Each color curve demonstrates two strong resonant maxima. The broader resonances are related to the absorption maximum at the interband transition frequency. Indeed, the position of the broader peaks coincides with the resonance in the continuous MCNG (black curve). The appearance of the second (narrow) peak in each curve is very different from the resonance in the case of the TM polarization. First, the narrow peak appears at lower frequency compared to the interband transition. This is clearly related to the spectral region of the TE GPPs (located below the interband transition, as shown in the bottom panel of Fig.5c). Second, the frequencies of the narrow resonance is almost independent upon the ribbon width. The inset in Fig. 5c showing the zoomed-in frequency region of the narrow resonances demonstrates the minor changes in the magnitude of the peak with change of $W$. The minor sensitivity of the frequency position of the narrow resonance to the ribbon width is related to a different nature of the resonance. Because the momentum of the TE GPPs is very close to the light line, they can easily leak out of the ribbon while reflecting from the edges, and therefore cannot build the Fabry-Perot resonances in individual ribbons. Instead, the TE GPPs in the grating constitute the Bloch modes and lead to the "collective" Bragg's resonances. In this case the resonance condition for the TE GPP refractive index is linked to the grating's period, $L$, (rather than to the ribbon width): $q_{pl} \approx \sin\theta + m\frac{\lambda}{L}$, where $m = \pm 1, ...$ is the diffraction order and $\theta$ is an incident angle. The dependence of the TE Bragg's resonances upon the angle of incidence and frequency is illustrated in more details in the Supplementary Note 3.

Due to the weak confinement, TE GPPs are far less sensitive to the dielectric environment of graphene. Indeed, in the inset of Fig.5c the absorption spectra by the free-standing array (dashed curves) are only slightly shifted from those by the hBN-encapsulated ribbon array (solid curves). Note that according to the dispersion curves shown in the bottom panel of Fig. 5c for the GPPs in the free-standing and hBN encapsulated graphene, the vertical confinement (scaling with by the inverse z-component of the k-vector $(1 - q_{pl}^2)^{-1/2}$) in the latter case is two times stronger. Nevertheless, as this confinement is still two orders of magnitude larger than the thickness of the considered hBN slab, the change in the dielectric environment does not considerably shift the resonances.

We thus demonstrate that MCNG micro-ribbon arrays are a powerful system for controlling the coupling between light and both TM and TE GPP modes. This enables the enhancement and the tailoring of THz absorption using applied magnetic fields or changing ribbon width and period.

**Discussion**

Summarizing, our work demonstrates that a single-layer CN graphene biased by an external magnetic field supports magnetically-tunable interband GPPs of both TM and TE polarizations (the latter possessing two orders of magnitude better confinement than conventional TE GPPs in a doped graphene). We proposed various realistic scenarios for the experimental observation of the GPP in MCNG: by means of the near-field optical microscopy (via s-SNOM) and far-field spectroscopy, corroborating both scenarios by the simulations. Importantly, we demonstrated that rather low values of magnetic fields ($B < 2\,T$) are fully suitable for the observations.

From a different perspective, we believe that our analysis of plasmons in a single-layer magnetized CN graphene can be relevant for understanding recently predicted collective excitations in more sophisticated structures, such as e.g. magic angle twisted bilayer graphene [13] or periodically strained single-layer CN graphene, and particularly, their puzzling topological states [11, 12, 14]. We foresee that the intriguing phenomena recently discovered in the single-layer CN graphene, as for instance the giant intrinsic photoresponse [11], colossal magneto-optical activity [10] and markedly reduced electronic noise [43], combined with the ability to support GPPs, make magnetized CN graphene promising for a new type of graphene-based magnetically controllable nanoplasmonic and optoelectronic devices, such as gas sensors or bio-sensors [43-45] or magnetically tunable plasmon-assistant photodetectors [46], among other.

# Figures

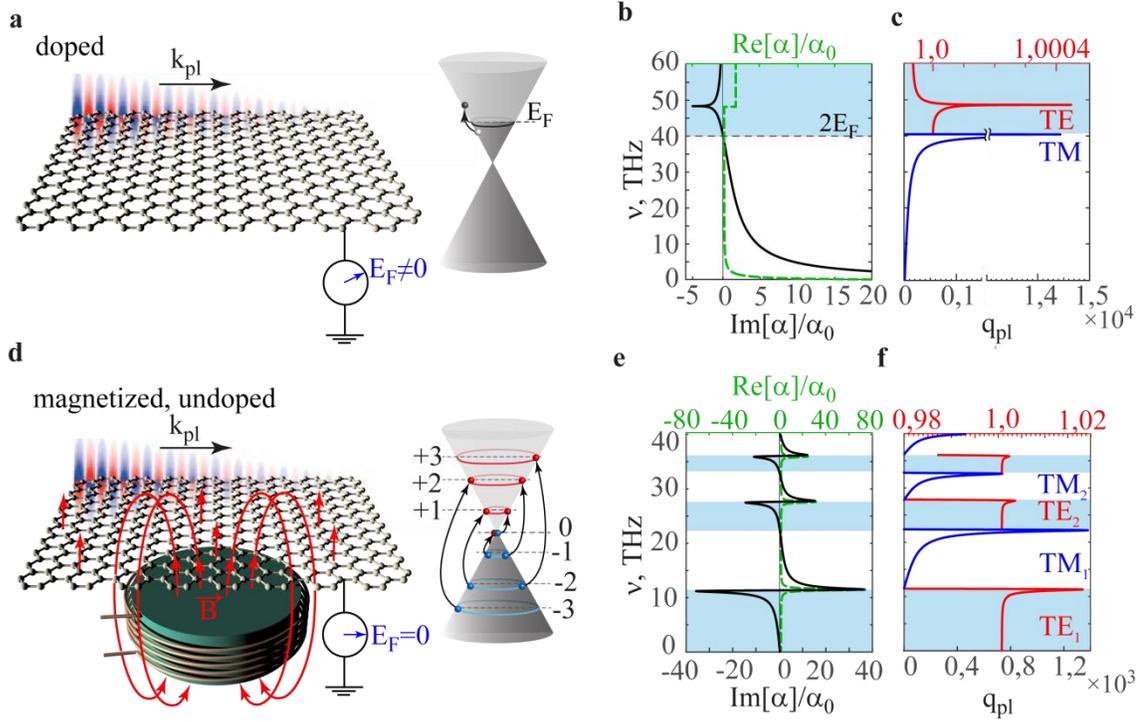

**Fig. 1**: The dispersion of GPPs in doped non-magnetized graphene and magnetized CNG. **a, d** Schematics for propagating GPPs. Schematics of optical transitions between the valence band and the conduction band. The electron filling is shown by the dark gray color. LLs are shown by blue and red lines in the Dirac cones. **b, e** Real and imaginary parts of the normalized conductivity (shown in units of the fine structure constant $\alpha_0 = 1/137$) as a function of frequency, $\nu$. The frequency bands, where $\mathrm{Im}[\alpha] > 0$ and $\mathrm{Im}[\alpha] < 0$, are highlighted by white and blue horizontal stripes, respectively. **c** The dispersion curves for GPPs in doped graphene calculated at $T = 0\,K$, $\tau = 1\,ps$, $E_F = 0.1\,eV$. **f** The dispersion curves for GPPs in MCNG calculated at $T = 0\,K$, $\tau = 1\,ps$, $B = 1.3\,T$

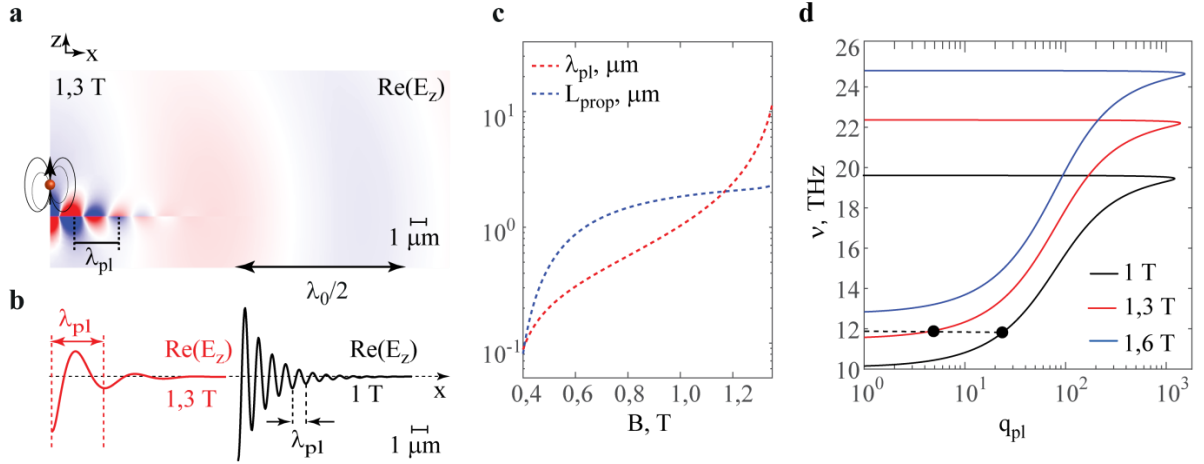

**Fig. 2**: Dependence of the properties of TM GPPs in MCNG upon the magnetic field. **a** Electric field snapshot for the TM GPP in MCNG, at $B = 1.3\,T$. **b** Real part of the z-component of electric field calculated for $B = 1.3\,T$ and $B = 1\,T$ (with wavelengths $\lambda_{p1} = 4.3\,\mu m$ and $\lambda_{p2} = 0.97\,\mu m$, respectively). **c** Propagation lengths and wavelengths of the TM GPP in MCNG as a function of the magnetic field. **d** Dispersion curves for TM GPP in MCNG calculated for different magnetic field. In **a,b,c** $\nu = 12\,THz$ ($\lambda_0 = 25\,\mu m$).

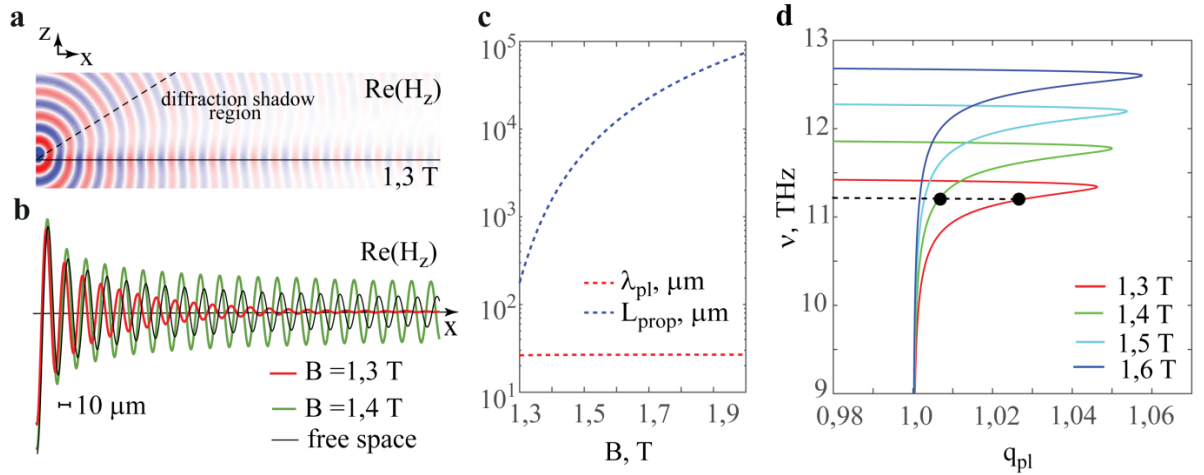

**Fig. 3**: Dependence of the properties of TE GPPs in MCNG on magnetic field. **a** Snapshot of the vertical component of the magnetic field of a TE GPP in MCNG field, at $B = 1.3\,T$. **b** Real part of the z-component of magnetic field calculated for $B = 1.3\,T$ and $B = 1.4\,T$, (the corresponding GPP wavelengths are $\lambda_{p1} = 26.32\,\mu m$ and $\lambda_{p2} = 26.66\,\mu m$, respectively). For comparison, the free-space wave oscillation (with the wavelength $\lambda_0$) is shown by the black curve. **c** Propagation lengths and wavelengths of the TM GPP in MCNG as a function of the magnetic field. **d** Dispersion curves for TE GPP in MCNGs, calculated for different magnetic fields. In **a,b,c** $\nu = 11.2\,THz$ ($\lambda_0 = 26.77\,\mu m$).

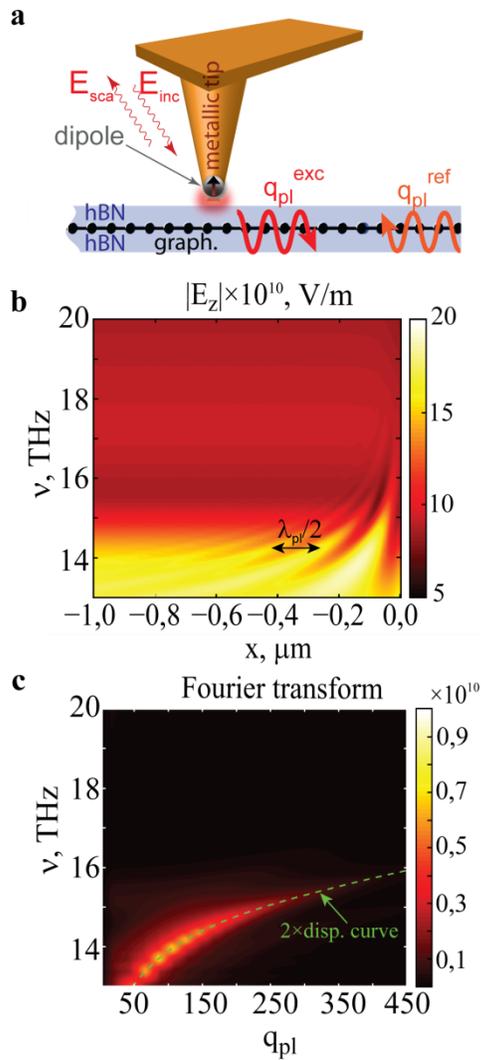

**Fig. 4**: Simulation of a s-SNOM experiment. **a** Schematics of the simulation model: the AFM tip approximated by an electric dipole source. The dipole is located $55\,nm$ above the CNG encapsulated into hBN (the thicknesses of the top and bottom layers are $5\,nm$ and $10\,nm$, respectively). **b** Simulated near-field image $|E_z(x,\nu)|$ at $B = 1.3\,T$. The field $E_z$ is taken below the dipole, $10\,nm$ above the graphene. **c.** The Fourier transform of $|E_z(x,\nu)|$. The bright feature matches twice the analytical dispersion of TM GPP in MCNG (shown by the dashed green curved).

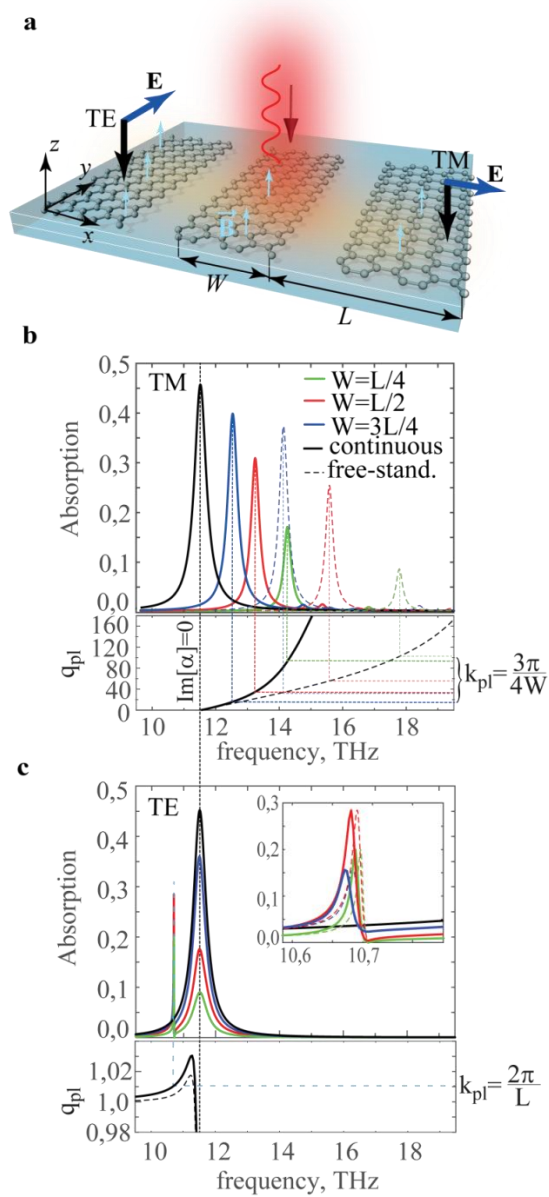

**Fig. 5.** Excitation of TM and TE GPP in MCNGs ribbon arrays. **a** Schematics of the structure. **b** Absorption spectra, $A(\nu)$, of the TM polarized light incident in hBN-encapsulated (solid curves) and free-standing (dashed curves) MCNG ribbon arrays with different ribbon widths. The thickness of both top and bottom hBN layers is $10\ nm$. The bottom panel demonstrates the dispersion of TM GPP in MCNGs. **c** The same as in **b**, but for the TE polarized light. The inset in the upper panel shows the zoom-in of the resonance region. The periods of the ribbon arrays are $L = 0.22\ \mu m$ and $L = 28\ \mu m$ for the TM and TE polarizations, respectively. In both cases, the magnetic field is $B = 1.3\ T$.

**Methods**

**First principle numerical simulation.** Full wave electromagnetic simulations were performed using the COMSOL software based on finite-element methods in frequency domain. The graphene layer was modeled as a surface current, implemented in the boundary conditions. In order to achieve convergence, the mesh element size in the vicinity of graphene was much smaller than the plasmon wavelength.

**Simulation of NF experiment.** In the simulation the tip was modeled by a vertical point dipole source (polarized along $z$-axis). We assume that the vertical component of the field below the dipole, $E_z$, approximates the scattered signal in a s-SNOM experiment [27]. We thus simulated the near-field profiles by recording the calculated $E_z$ as a function of the dipole position, $x$ (due to the translational symmetry of the problem along the $y$-axis, $E_z$ does not depend upon $y$), and dipole frequency. The field $E_z$ is taken 5 $nm$ above the hBN, below the dipole. Therefore, we achieve NF in the space–frequency domain $(x, \omega)$ and, we perform its Fourier transform to obtain the dispersion diagram of GPP in MCNG.

**Data availability**

The data supporting the findings of this study are included in the main text and in the Supplementary Information files, and are also available from the corresponding authors upon reasonable request.

**Acknowledgements**

We acknowledge funding from Spain's MINECO under Grant No. MAT2017-88358-C3 and funding from the European Union Seventh Framework Programme under grant agreement no.785219 Graphene Flagship for Core2. L.M.-M and T.M.S acknowledge Aragon Government through project Q-MAD.

**Author contributions**

A.B.K., A.Y.N. and L.M.-M conceived the work. T.M.S., A.Y.N. and L.M.-M performed numerical and analytical calculations. T.M.S and A.Y.N. wrote the manuscript with the input from all co-authors. All the authors contributed to the discussion of the results.

Corresponding Authors

[*] Email: lmm@unizar.es

[*] Email: Alexey.KuzMenko@unige.ch

[*] Email: tetiana@unizar.es

**Competing interests**

The author(s) declare no competing interests.

**Additional information**

**Supplementary Information** accompanies this paper at https://

**Interband plasmon polaritons in magnetized charge-neutral graphene**

T. M. Slipchenko[1,2*], J.-M. Poumirol[3,4], A. B. Kuzmenko[3*], A. Yu. Nikitin[5, 6], L. Martín-Moreno[1,2*]

[1] Instituto de Nanociencia y Materiales de Aragón (INMA)

CSIC-Universidad de Zaragoza, Zaragoza 50009, Spain

[2] Departamento de Física de la Materia Condensada,

Universidad de Zaragoza, Zaragoza 50009, Spain

[3] Department of Quantum Matter Physics, University of Geneva,

[4] CEMES, University of Toulouse, CNRS, Toulouse 31000, France

[5] IKERBASQUE Basque Foundation for Science, 48013 Bilbao, Spain.

[6] Donostia International Physics Center (DIPC), 20018 Donostia-San Sebastin, Spain

## Supplementary Note 1

In Supplementary Note 1 we demonstrate how to change dispersion of GPP in MCNG with increase of both graphene doping and temperature.

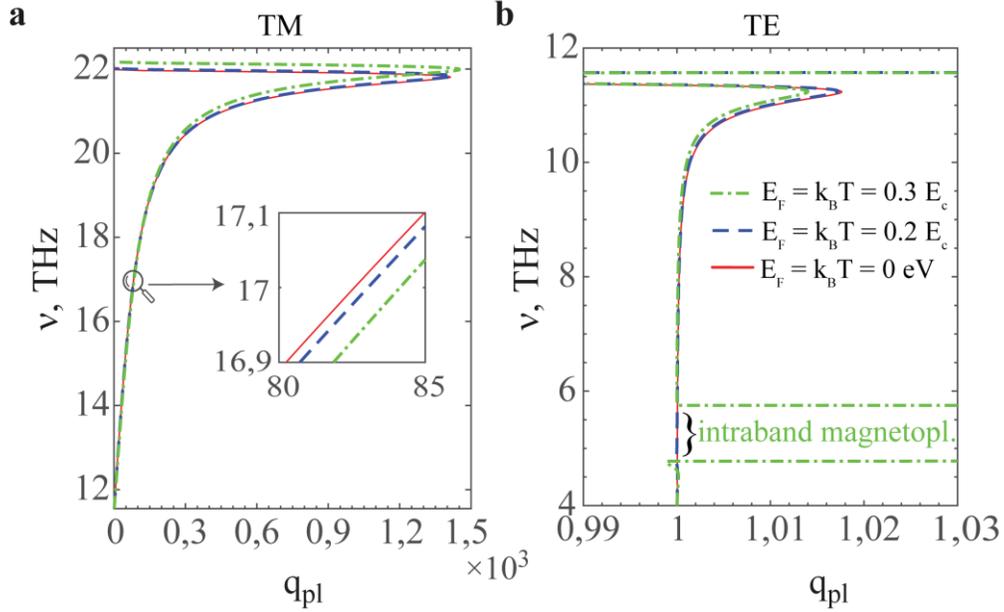

**Fig. S1**: The dispersion of GPPs in MCNG for non-zero values of both Fermi energy and temperature calculated for both **a** TM and **b** TE mode polarizations. The dispersion curves calculated for $\tau = 1\ ps$, $v_F = 1.15 \times 10^6 ms^{-1}$, $B = 1.3\ T$ ($E_c = 0.048\ eV$). The inset in the panel **a** shows the zoom-in of the region of the dispersion curves. The abrupt growth of $q_{pl}$ seen in the frequency window between 4.7 THz and 6 THz in the green dashed-doted dispersion curve in the panel **b** correspond to the dispersion of the intraband GPP appeared due to electron transition between $1^{st}$ and $2^{nd}$ LLs.

As seen in Fig. S1, the dispersions of GPP undergo negligible changes with an increase of both temperature, $k_B T$, and Fermi energy, $E_F$, up to $0.2\ E_c(B)$. Simultaneously rising $E_F$ and $k_B T$ from zero up to $0.2 E_c(B)$ at frequency 17 THz the dispersion shifts only by $\Delta q_{pl} \approx 0.5$. However, increasing further $k_B T$ and $E_F$ up to $0.3\ E_c(B)$ the dispersions of GPP starts to differ notably from one of the GPP in MCNG at $T = 0\ K$, especially for TE modes. For example, at frequency 17 THz the TM mode dispersion shifts only by $\Delta q_{pl} \approx 1.6$, but the TE mode dispersion demonstrates abrupt growth of $q_{pl}$ in the frequency window between 4.7 THz and 6 THz.

The fact is that with the simultaneous increase of $E_F$ and $k_B T$ up to $0.3\ E_c(B)$ the occupancy of the zeroth LL rises from 0.5 till 0.73 and the occupancies of higher LLs become to differ from zero. The thermal fluctuations thus can provide the charge transition between occupied LLs within the same band at the discrete frequencies $v_n = E_c\left(\sqrt{|n+1|} - \sqrt{|n|}\right)/h$. At these frequencies the imaginary part of graphene conductivity changes its sign from negative to positive what makes possible the existence of intraband magnetoplasmons in the graphene. It is also worth to note that the non-diagonal conductivity of slightly doped graphene, $\sigma_{xy}$, is not zero anymore. Due to this fact, GPP modes of magnetized graphene are hybridized containing both TM and TE polarization components at the same moment.

## Supplementary Note 2

In Supplementary Note 2 we demonstrate how to change the figure of merits of GPP in MCNG with increase of applied magnetic field.

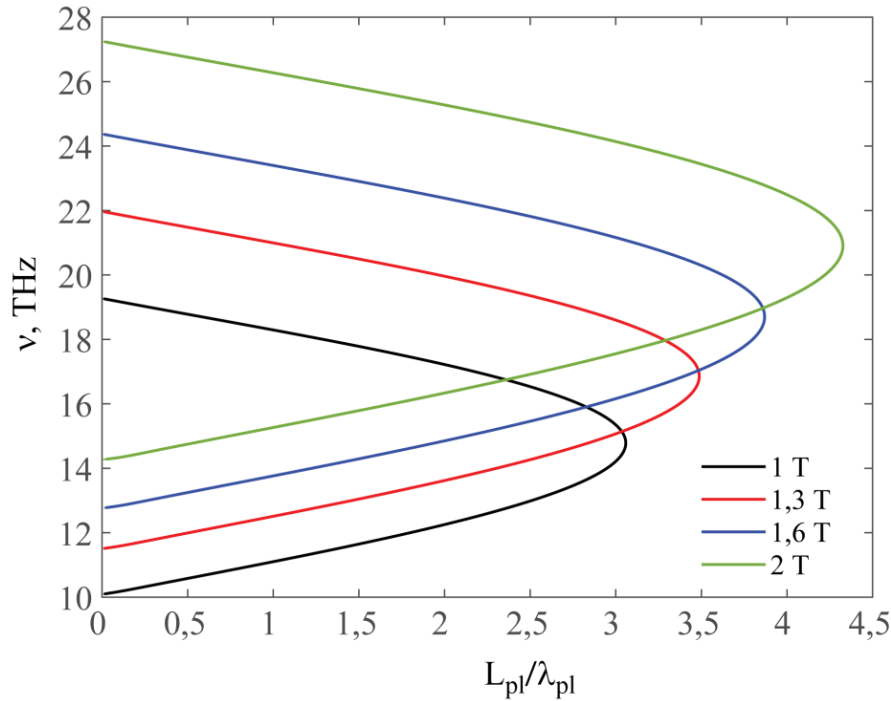

**Fig. 2S**: Figure of merit of TM GPPs in MCNG calculated for different magnetic fields within the first frequency band. The figure of merits calculated for MCNG at $\tau = 1\ ps$, $v_F = 1.15 \times 10^6 ms^{-1}$

Remind that figure of merits (FOM) is defined as the ratio between GPP propagation length, $L_{pl}$, and the GPP wavelength, $\lambda_{pl}$, FOM $= L/\lambda_{pl} = \text{Im}(\alpha)/2\pi\text{Re}(\alpha)$.

Fig. 2S shows FOM of TM GPPs in MCNG upon different magnetic fields within the first frequency band. For all values of the magnetic fields FOM spectrum presents a maximum. This maximum appears in the middle of the frequency band, at that in the begging and at the end of the frequency band FOM is equal to zero due to the vanishing of $\text{Im}(\alpha)$ at these frequencies.

## Supplementary Note 3

In Supplementary Note 3 we illustrate the dependence of the TE Bragg's resonances appeared in the upon MCNG ribbon array on the angle of incidence and frequency.

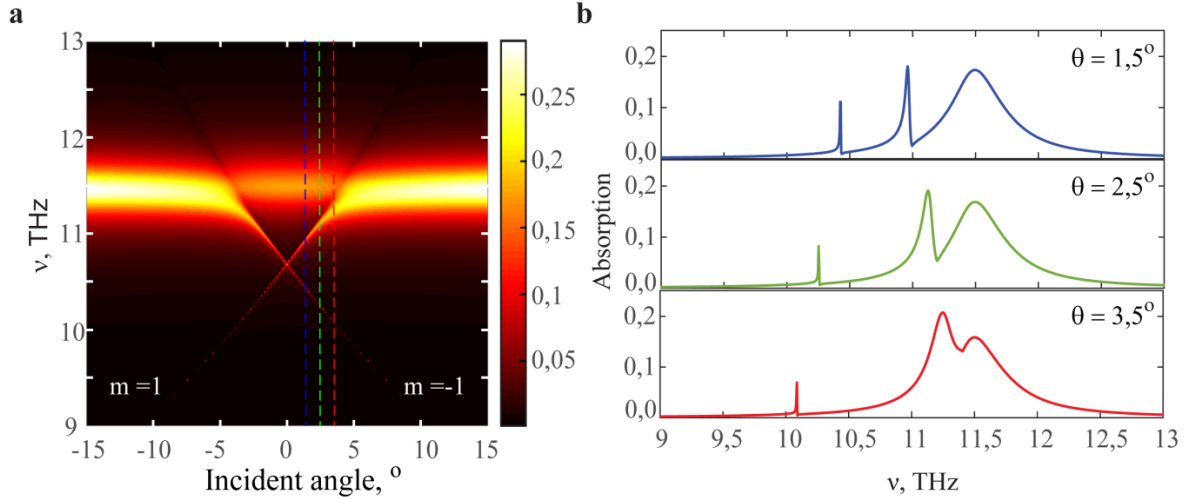

**Fig. 3S.** Angle dependence of absorption of MCNG ribbon array. **a** Absorption as a function of angle of incidence and frequency for ribbon array with the ribbon width equaled to $W = L/2$. **b** Absorption spectra calculated for the incident angles denoted by the dashed lines in the panel a. The period is $L = 28\,\mu m$ Magnetic field amplitude in both cases is $B = 1.3\,T$ Other parameters are $\tau = 1\,ps$, $v_F = 1.15 \times 10^6 ms^{-1}$

In Fig. 3S in panel a we present the absorption as a function of angle of incidence and frequency for ribbon array with the ribbon width equaled to one half of period, $W = L/2$. We see that absorption resonances are strongly depends on the incident angle. It can be explained by Bragg's nature of the resonances which can be described by the condition: $q_{pl} \approx \sin\theta + m\frac{\lambda}{L}$, where $m = \pm 1, ...$ is the diffraction order, $\theta$ is an incident angle. Thin red ridges seen in the color plot in the frequencies below $11\,THz$ correspond to the excitation of TE GPP in the 1st and -1st diffraction order. At frequencies close to the interband transition frequency, $v_n \approx 11.5\,THz$, TE plasmon resonance and the resonance due to the interband charge transition couple. This coupling can be seen better in the panel b of Fig. 3S where the absorption spectra calculated for different incident angles are presented. In fact, with incident angle grows the absorption resonance due to the TE GPP excitation in the 1st diffraction order approach to the absorption resonance at the interband transition frequency.